\documentclass[a4paper,11pt]{article}
\usepackage{pos}

\makeatletter
\renewenvironment{thebibliography}[1]%  %UNFORTUNATELY MODIFIED..
     {\bgroup\raggedright\small\section*{\refname
        \@mkboth{\MakeUppercase\refname}{\MakeUppercase\refname}}%
      \list{\name{bib\@arabic\c@enumiv}% HOPE!
            \@biblabel{\@arabic\c@enumiv}}%
           {\settowidth\labelwidth{\@biblabel{#1}}%
            \setlength\itemsep{0pt}
            \setlength\parskip{0pt}
            \setlength\parsep{0pt}
            \setlength\partopsep{0pt}
            \leftmargin\labelwidth
            \advance\leftmargin\labelsep
            \@openbib@code
            \usecounter{enumiv}%
            \let\p@enumiv\@empty
            \renewcommand\theenumiv{\@arabic\c@enumiv}}%
      \sloppy\clubpenalty4000\widowpenalty4000%
      \sfcode`\.\@m}
     {\def\@noitemerr
       {\@latex@warning{Empty `thebibliography' environment}}%
      \endlist\egroup}
\makeatother

\title{Subatomic Heroes}
\author[a]{Anshika\,Bansal}
\author[a]{Guido\,Bell}
\author[a]{Aritra\,Biswas}
\author[a]{Diptaparna\,Biswas}
\author[a]{Anastasia\,Boushmelev}
\author[b]{Carsten\,Busse}
\author[a]{Markus\,Cristinziani}
\author[a]{Carmen\,Diez\,Pardos}
\author[a]{Qader\,Dorosti}
\author[a]{Sebastian\,Edelmann}
\author[a]{Thorsten\,Feldmann}
\author[a]{Ivor\,Fleck}
\author[a]{Jan\,Hahn}
\author[a]{Dennis\,Horstmann}
\author[a]{Tobias\,Huber}
\author[a]{Jack\,Jenkins}
\author[a]{Wolfgang\,Kilian}
\author[a]{Danny\,Koschwitz}
\author[a]{Nils Krengel}
\author[a]{Martin\,Lang}
\author[a]{Björn\,Lange}
\author[a]{Alexander\,Lenz}
\author[a]{Eleftheria\,Malami}
\author[a]{Thomas\,Mannel}
\author[a]{Ilija\,Milutin}
\author[a]{Ali\,Mohamed}
\author[a]{Denise\,Müller}
\author[a]{Jakob\,Müller}
\author[a]{Marcus\,Niechciol}
\author[a]{Jaime\,del\,Palacio\,Lirola}
\author[a]{Chiara\,Papior}
\author[a]{Arwen\,Pieck}
\author[c]{Thomas\,Reppel}
\author[a]{Markus\,Risse}
\author[a]{Elisabeth\,Schopf}
\author[a]{Noah\,Siegemund}
\author[a]{Tobias\,Striegl}
\author[a]{Gilberto\,Tetlalmatzi-Xolocotzi}
\author[a]{Tom\,Tong}
\author[a]{Jette\,Vedder}
\author[a]{Katharina\,Voß}
\author[a]{Daniel\,Vladimirov}
\author[a]{Wolfgang\,Walkowiak}
\author*[a]{Oliver\,Witzel}
\author[a]{Zachary\,Wüthrich}
\author[b]{Christof\,Wunderlich}
\author[a]{Michael\,Ziolkowski}

\affiliation[a]{Center for Particle Physics Siegen (CPPS)\\
  Universität Siegen, Walter-Flex-Straße 3, 57072 Siegen, Germany}
  
\affiliation[b]{Physik Department\\
  Universität Siegen, Walter-Flex-Straße 3, 57072 Siegen, Germany}

\affiliation[c]{Dekant und Fakultät IV MINT Öffentlichkeitsarbeit\\
  Universität Siegen, Hölderlinstraße 3, 57076 Siegen, Germany}

\emailAdd{physik-outreach@uni-siegen.de}

\abstract{Sharing the amazing achievements of the (particle) physics world with the general public is at the heart of the mission of the Subatomic Heroes, based at the University of Siegen, Germany. 
Originally this started out as an endeavor of theoretical particle physics, now we are steadily spreading out to cover and include more branches of physics and science.
Our activities range from merging art with public physics lectures via marvelous artistic performances at the local theater, over dedicated events for high-school students, to our Subatomic Heroes channel on Instagram and TikTok where you may also find out when and where our famous ``hadronic ice-cream'' will be served next! So  follow us on \url{https://www.instagram.com/subatomic_heroes} and \url{https://www.tiktok.com/@subatomic_heroes}.}

\FullConference{The European Physical Society Conference on High Energy Physics (EPS-HEP2025)\\
7-11 July 2025\\
Marseille, France\\}

\begin{document}
\maketitle

\begin{figure}[t]
  \centering
  \includegraphics[width=4cm]{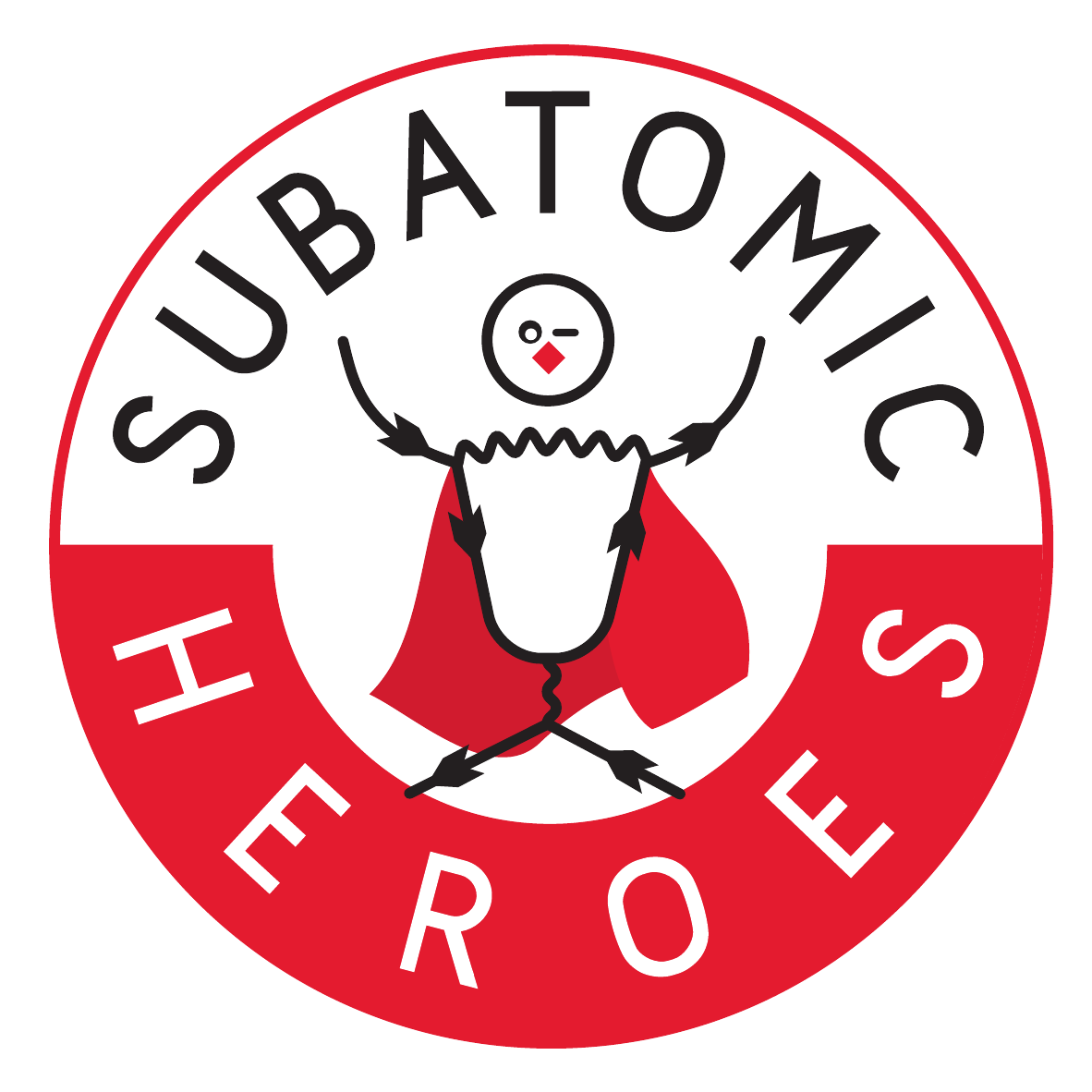}
  \caption{The logo of the Subatomic Heroes. The black contours of our superhero depicts actually a 
  peculiar  elementary particle process: the decay of e.g.~a $b$-quark into a $s$-quark and a muon and anti-muon pair - a so-called electro-weak penguin diagram. Interestingly we find for more than 10 years some discrepancies between experimental measurements and theoretical expectations of these process. Maybe our superhero lies at the heart of some future exciting discoveries in physics.}
  \label{Fig.logo}
\end{figure}

\section{Motivation}
At the 11\textsuperscript{th} International Workshop on Charm Physics (CHARM 2023) in Siegen, Germany, \cite{Charm2023} the Subatomic Heroes, see the logo in Fig.~\ref{Fig.logo},  entered for the first time the big, international stage and have ever since pursued their threefold mission:
\begin{enumerate}
\item \textbf{Attract students to study physics in Siegen.} Siegen is a town of about 100,000 people located centrally in Germany and prospective students have many opportunities to study physics in nearby cities or metropolis regions. So why study physics in Siegen? 
We have an amazing staff-student ratio, which creates an atmosphere where each student can easily talk to their professors and students know their classmates.
Moreover, the Department of Physics features exceptionally strong research groups e.g.~in particle physics or in  quantum optics with some of the world-leading experts in their field. This was recently officially certified with the award of the  Deutsche Forschungsgemeinschaf (DFG) excellence cluster {\it Color meets Flavor} \cite{CmF}.
\item \textbf{Spark pupils' interest in (particle) physics and science.} The joy of physics and our curiosity to understand Nature is driving our research and exciting us every day. Sharing this excitement with the young generations is a central part of our mission. We are keen on inspiring prospective students of tomorrow and we reach out to all genders and all backgrounds, starting already in primary school.
\item \textbf{Education for the general public.}
In particular the experiences during the epidemic have shown that the need for support and an understanding of the necessity of fundamental research by the general public cannot be overestimated. This is of course impeded by abstract and complicated concepts in science.
Therefore, we put a big emphasis on communicating to the general public and developing new ideas and concepts for that.
\end{enumerate}
In the following we will give an overview of the different adventures the Subatomic Heroes have undertaken and group them according to the three aspects of our mission.

\section{Attract Students}
\begin{figure}[tb]
  \begin{minipage}{0.6\textwidth}
  \includegraphics[height=0.3\textheight]{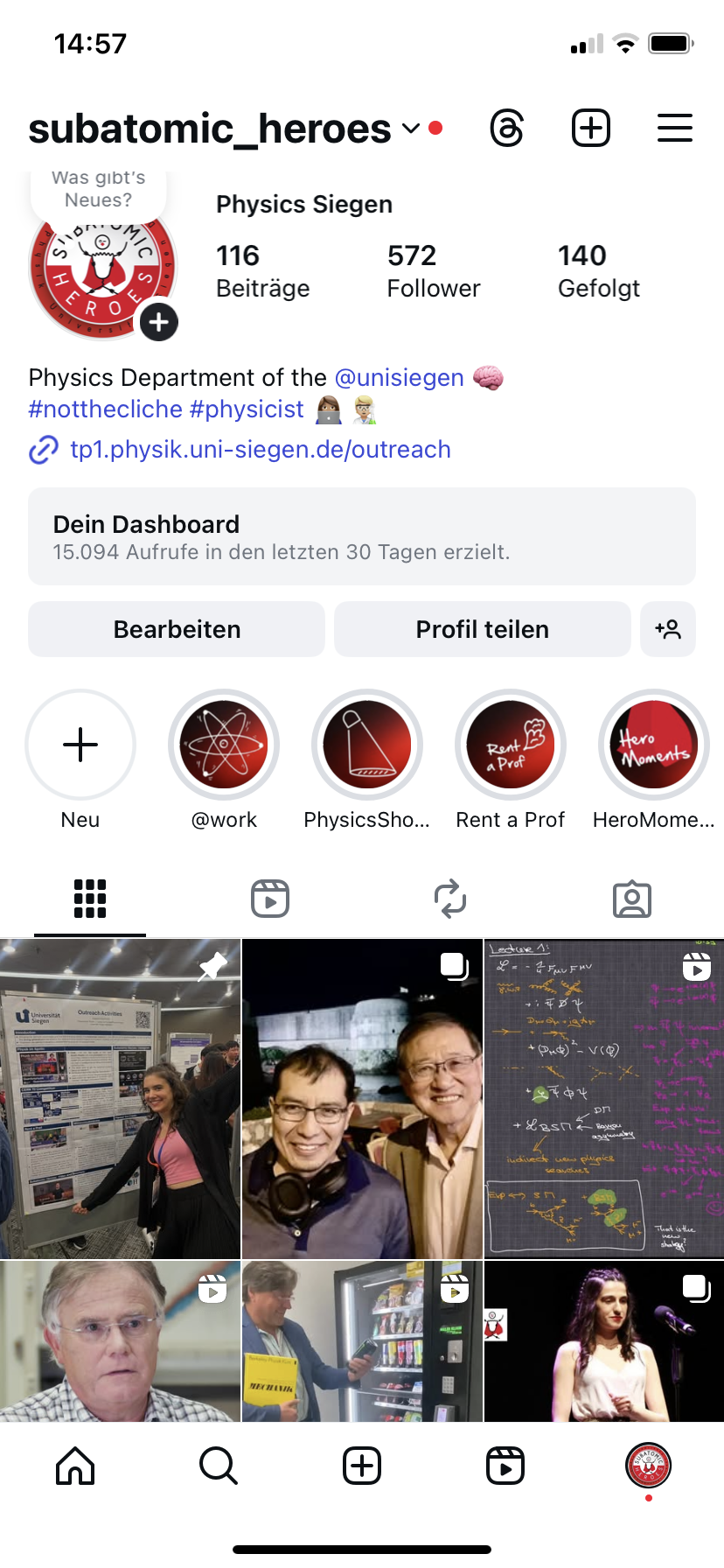}\hfill
  \includegraphics[height=0.3\textheight]{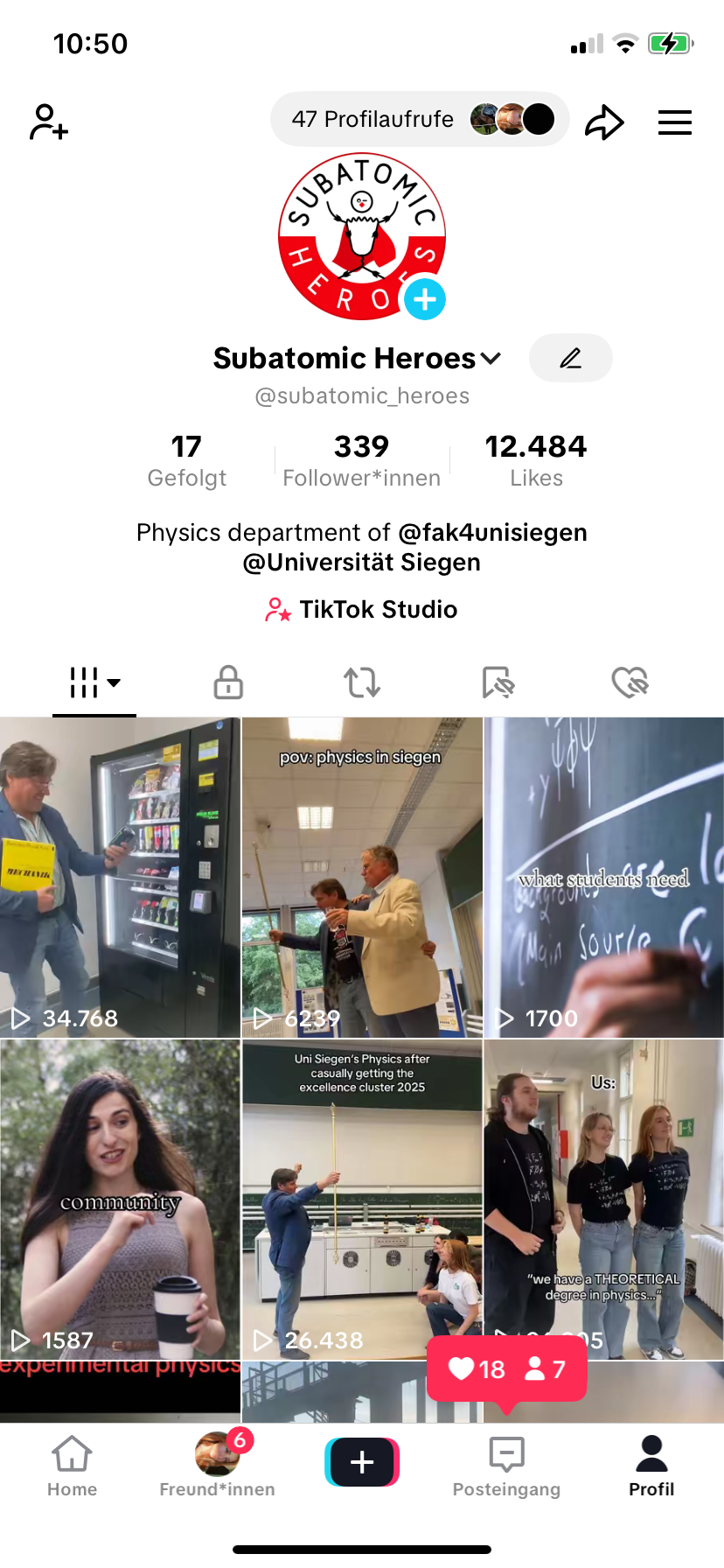}
  \end{minipage}\hfill
  \begin{minipage}{0.32\textwidth}
  \includegraphics[width=\textwidth]{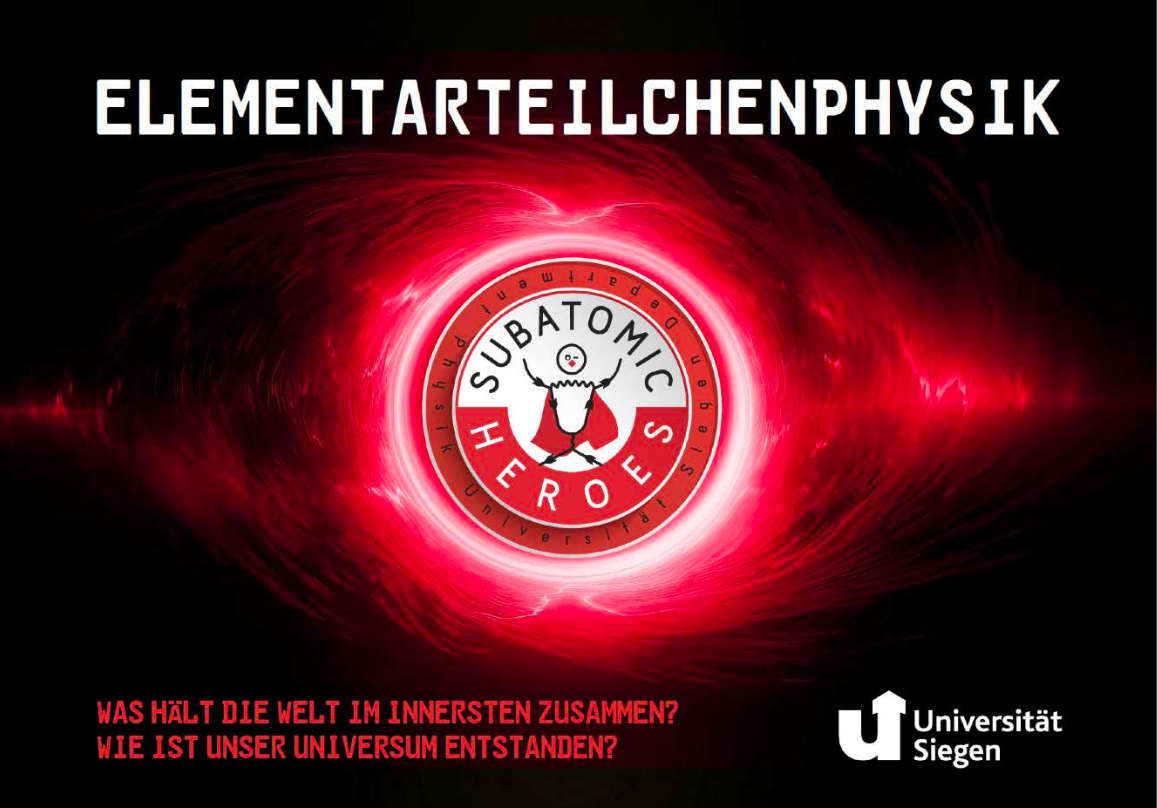}\\[2mm]
  \includegraphics[width=\textwidth]{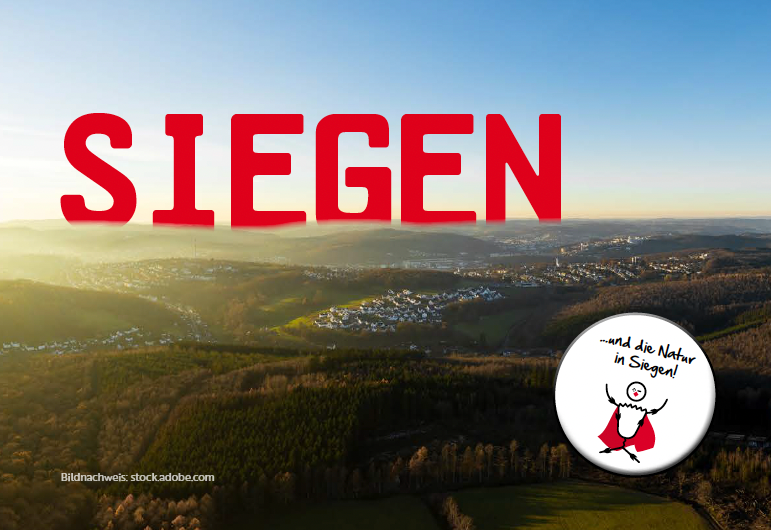}
  \end{minipage}
  \caption{Screenshots of our Instagram channel (left), TikTok channel (center), and sample pages from our brochure ``Elementarteilchenphysik'' (right).}
  \label{Fig.Students}
\end{figure}

With our social media channels \cite{Instagram,TikTok}
we convey the messages: studying physics in Siegen is fun and incredibly interesting with many exciting events and activities ongoing. The channels are predominantly run by PhD students and post-docs (Instagram for the young generations) and by Bachelor-students (TikTok for the very young generations). 
They report on physics and life in academia, use our channel to broadcast academic and outreach events in Siegen, or discuss physics highlights. When traveling, we also report from international conferences and workshops. Screenshots of our Instagram and TikTok channel are shown in Fig.~\ref{Fig.Students}.

In addition we prepared a brochure 
\href{https://tp1.physik.uni-siegen.de/wp-content/uploads/sites/6/2024/09/broschure_UNI-compressed-1.pdf}{``Elementarteilchenphysik'' (elementary particle physics)},
where we give a layman introduction to elementary particle physics and the Standard Model. We discuss how we obtain our knowledge, what open questions are driving us, and which expertise the research groups in Siegen have and we point out open questions in our current understanding of the world. Of course we also report on studying physics in Siegen and who the Subatomic Heroes are. Two sample pages are shown in the right of Fig.~\ref{Fig.Students}. The brochure is accompanied by a \href{https://tp1.physik.uni-siegen.de/wp-content/uploads/sites/6/2025/10/subatomic_heroes_englische_untertitel.mp4}{video} featuring the Subatomic Heroes and their mission --- please have a look! 
  
Furthermore, we regularly contribute to podcasts: Strange \& Charming \cite{StrangeCharming} was initiated by Robert V.~Harlander (RWTH Aachen University) as an outreach activity of our Collaborative Research Center (CRC TRR 257) \cite{P3H}. The concept is that physicists talk about their experiences and discuss questions like: Why did they study physics?  Which obstacles did they overcome? What is ``the daily life'' of a physicist? The different recordings are produced by CRC members in English or German. Similarly we have contributed episodes of the University of Siegen science podcast ``Spark!'' \cite{Spark}, where topics like {\it particle physics} or {\it cosmic rays}
are discussed in German.  

\section{Sparking Pupils' Interest}

\begin{figure}[tb]
  \includegraphics[height=0.13\textheight]{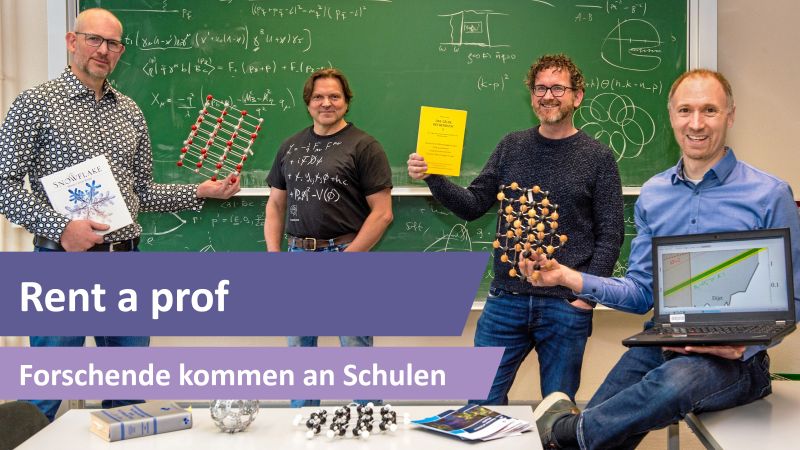}\hfill
  \includegraphics[height=0.13\textheight]{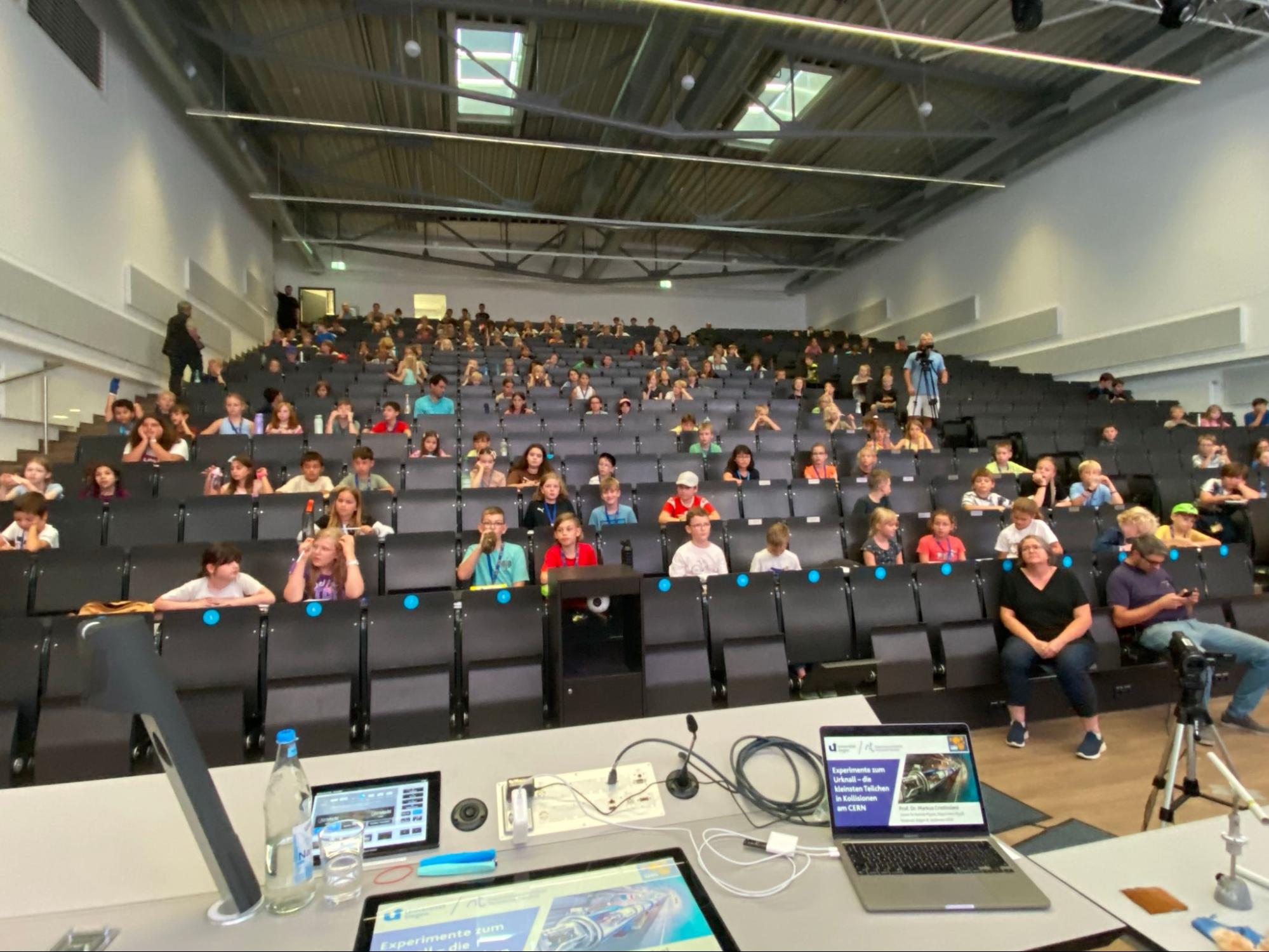}\hfill
  \includegraphics[height=0.13\textheight]{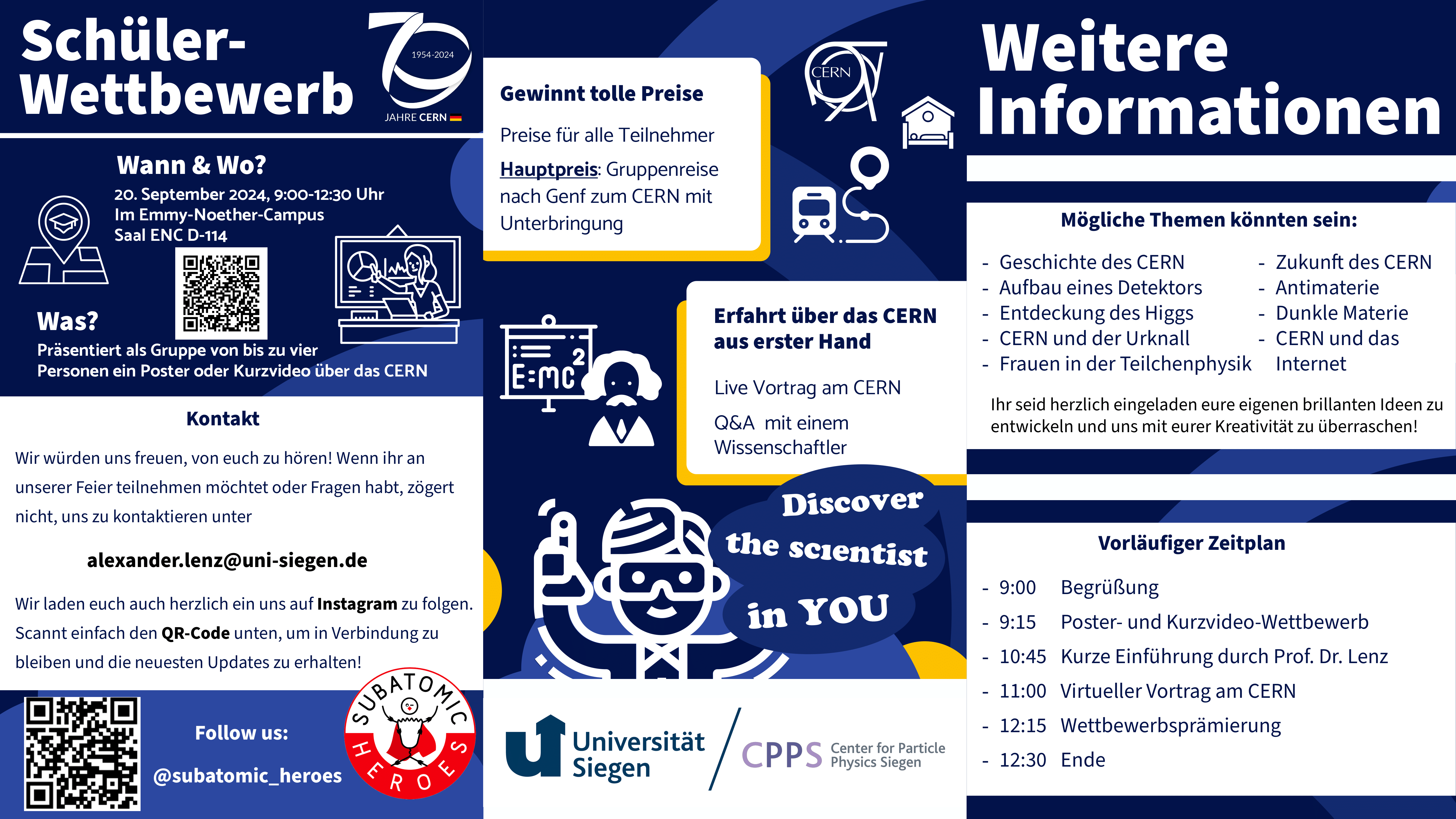}    
  \caption{Left: ``Rent a Prof'' program organized by our Faculty IV, University of Siegen, center: ``University for Children'', and right: flyer for our high school student competition celebrating CERN's 70\textsuperscript{th} anniversary.}
  \label{Fig.Pupils}
\end{figure}

To spark the interest of pupils in physics we run different programs targeting both primary as well as secondary school students and teachers. The program ``Rent a prof'' \cite{RentProf}, organized centrally for our Faculty IV  by Thomas Reppel, is intended for teachers from schools in the vicinity of Siegen to contact us in order to arrange special lessons (cf.~Fig.~\ref{Fig.Pupils}). University professors then visit the school and take over a lesson with topics like {\it What is the smallest thing in the Universe?} or  {\it Special Relativity} or  {\it Status Quo of Particle Physics}
--- including discussion sessions, which are particularly successful in primary schools. 

Moreover, we regularly offer weekly \href{https://tp1.physik.uni-siegen.de/mittwochsakademie/}{lecture courses} for talented and interested pupils, covering topics like {\it theoretical mechanics}, {\it quantum mechanics} or {\it quantum computing}.
In addition many high school students do their mandatory two-week internship at our department. Students at the age of about 15 years visit different working groups and mostly spend two days with each group. They get a chance to experience life at the university, setup an experiment from a lab course, attend a lecture, or a tutorial. Typically we organize a day in four 2h-slots and some slots are dedicated to touch base with them on different topics in math or physics.  The students benefit from private tutoring sessions, e.g., on academic careers, complex numbers, Monte Carlo integration, using computer algebra systems, or programming in general. 

Joining the community wide efforts to bring particle physics to high school students we are also participating in CERN Masterclasses \cite{Masterclass} organized in Germany by the Netzwerk Teilchenwelt \cite{TeilchenWelt}, where students experience analyzing real data from CERN experiments and rediscover e.g.~the Higgs boson. Further we  invite once per year the final classes of our nearby high schools (pupils age of 15+ years) to visit our Department of Physics and enjoy a day with lectures, visits to laboratories and  experimental physics demonstrations. Likewise we reach out to very young pupils (age 8-12 years) as part of the initiative ``Kinderuni'' (University for Children) \cite{KinderUni} see central panel of Fig.~\ref{Fig.Pupils}.
%where we bring our equipment to the downtown campus and present a lecture with experiments on the level of the kids. 

One of the highlights in the past year was the celebration of  CERN's 70\textsuperscript{th} anniversary.
%So we discussed how we can we best reach out to %students and pupils in our area to share the great %spirit of CERN and have a special event where they can %also actively contribute. In the end, 
We initiated a contest (see flyer on the right of Fig.~\ref{Fig.Pupils}) for groups of pupils to create a short video or presentation related to CERN, its history and future, famous scientists, etc. 
On the competition day we first screened all the contributions and enjoyed afterwards a virtual tour at CERN. The short-videos have shown an amazing level of creativity and professionalism --- please enjoy the winning videos yourself \cite{CERN70}.
Incentive for the participating pupils was the chance to win a 3-day visit to CERN  and the winning groups visited CERN in January 2025.

\section{General Public}
\begin{figure}[tb]
  \includegraphics[width=0.46\textwidth]{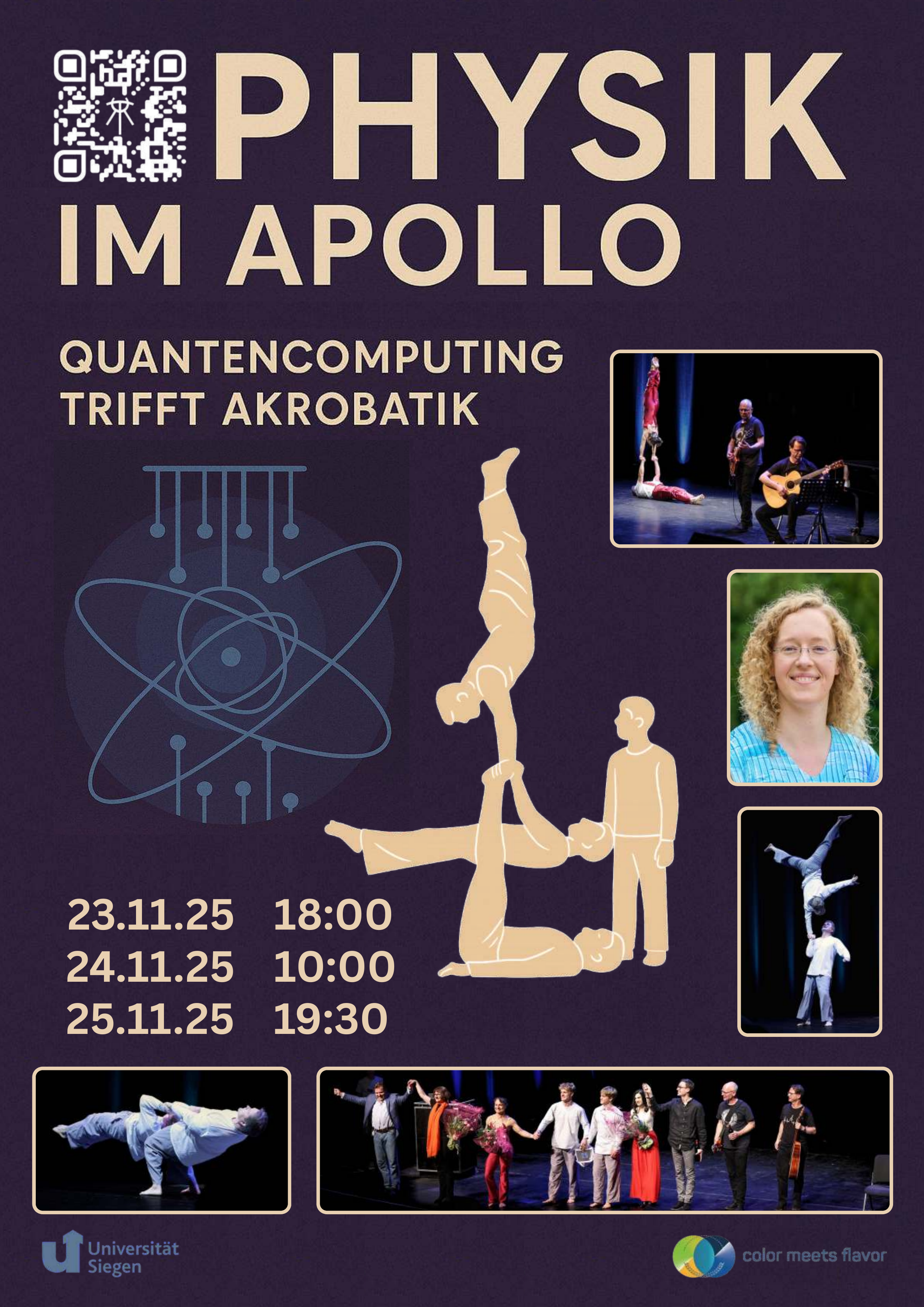}\hfill
  \includegraphics[width=0.46\textwidth]{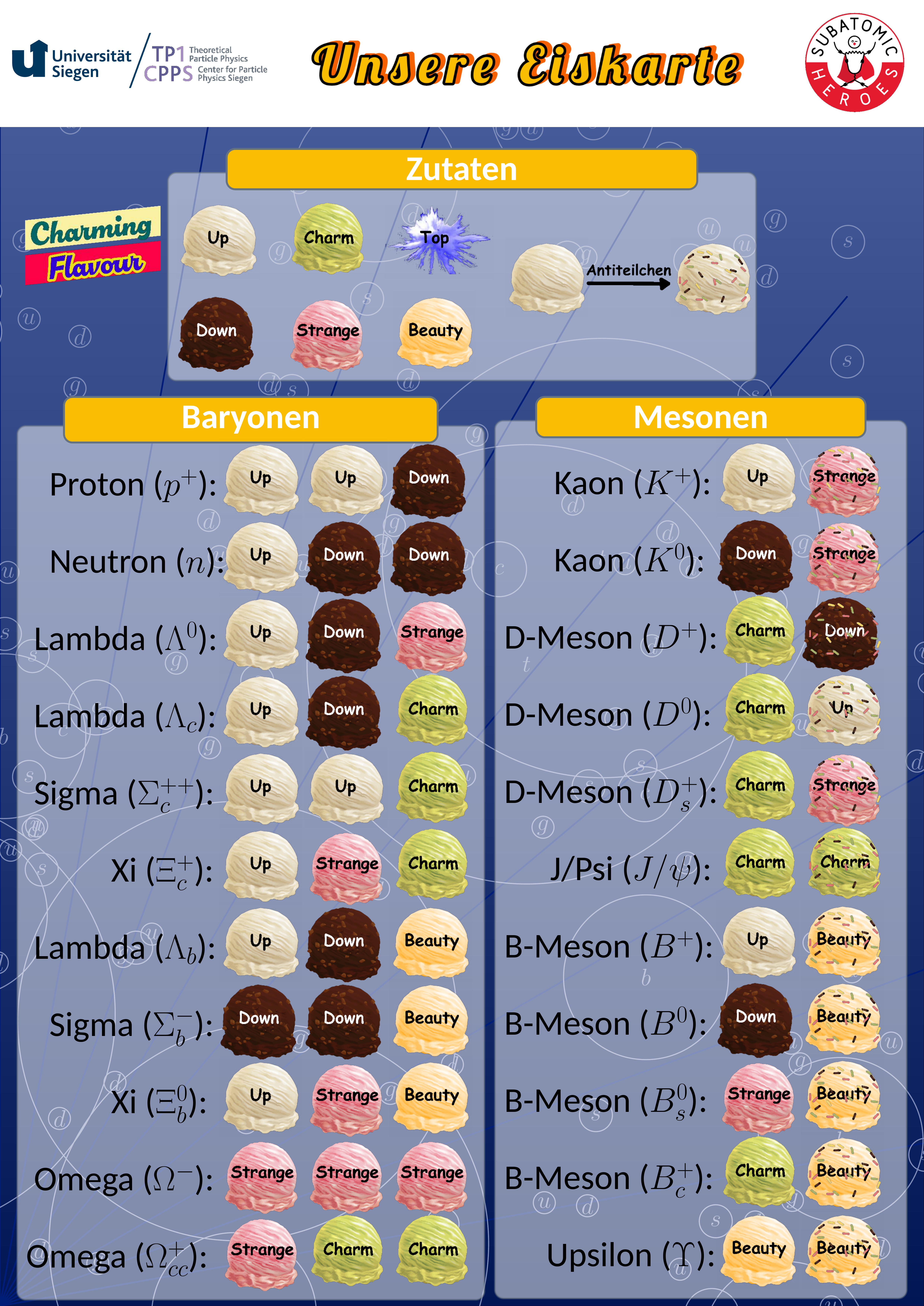}
  \caption{The left panel shows our ``Physik im Apollo'' 2025 poster, the right panel our hadronic ice-cream menu.}
  \label{Fig.Public} \vspace{-2mm}
\end{figure}

Reaching out to the general public can benefit from some additional highlights which trigger the public's curiosity or interest. In the past years, we have successfully explored different ideas. 
\\
``Physik im Apollo'' puts physics literally on  stage at the local Apollo theater in Siegen, 
which features 550 seats. Our concept is to educate 
the general public on abstract and difficult topics like {\it Dark Matter in the Universe} or {\it Quantum Computing} by interweaving popular science talks 
with performing arts, in particular acrobatics and music. 
An additional feature of our concept is that every artist  on stage is either physicist or has a very close connection to physics. 
This fall we will run ``Physik im Apollo'' for the third time and after sold-out shows in 2023 and 2024 we are again looking forward to a major highlight at the Apollo Theater Siegen. The poster is shown in Fig.~\ref{Fig.Public}.
The above mentioned  \href{https://tp1.physik.uni-siegen.de/mittwochsakademie/}{lecture courses} for  pupils, in {\it theoretical mechanics}, {\it quantum mechanics} or {\it quantum computing}
are also open for the general public under the lable
of \href{https://www.uni-siegen.de/hdw/formate/miak-root/semesterprogramm.html}{Mittwochsakademie}. 

To trigger curiosity at outdoor events like the open-day of the University, the ``Urknall Unterwegs'' (big bang on the road) organized by the Netzwerk Teilchenwelt provides a colorful attraction. 
It consists of different blow-up modules with the highlight a tall tunnel explaining the evolution of the Universe and the connection to particle physics. 
It also comes with activities and games for children of different age like a particle twister, creating badges with the animals of DESY's Teilchenzoo \cite{TeilchenZoo}. 
In addition we complement the exhibition with local demonstration experiments such as a mechanical scattering experiment or the cosmic can to visualize cosmic particles. We also provide stick-on tattoos with the Standard Model Lagrangian or Feynman diagrams. 
At the open-day of the University this year, we tried out a new idea to motivate guests to visit different booths of our Faculty IV (natural sciences and engineering) by jointly organizing a ``MINT-Rally''. 
MINT is a German acronym referring to mathematics, computer science, natural sciences and technology and the idea of the rally is that guests collect stamps by visiting the different booths in order to participate in a prize drawing.

One of our most favorite highlights is the ``hadronic ice-cream'' which has the amazing feature that the general public approaches us to learn more about particle physics. 
The idea is simple: we hand out free ice-cream and each scoop of ice-cream represents a different quark flavor. 
In addition, we use sprinkles to turn a quark into its anti-quark. 
\textbf{But} quarks occur only as confined hadrons and isolated quarks cannot be observed in Nature. 
Thus instead of ordering a scoop of ice-cream, our guests need to order mesons or baryons! 
Unfortunately, the lifetime of the top quark is too short to form flavored bound states. So up, down, charm, strange, and bottom form our hadronic ice-cream menu shown in Fig.~\ref{Fig.Public}.
As every good ice-cream parlor we also have a secret, specialty menu and welcome any order for a particle you find in the PDG \cite{PDG}.

\section{Summary}
The Subatomic Heroes love physics and are happy to share their enthusiasm with others. Using creativity and joy, we engage with all generations and hope this brief summary of their activities may inspire you to do the same. Please follow us on Instagram or TikTok and check-out our \href{https://tp1.physik.uni-siegen.de/outreach/}{outreach webpage} \cite{Website}. Of course you are also very welcome to drop by for our next round of ``hadronic ice-cream'' or the next ``Physik im Apollo'' event at the Apollo theater Siegen.

\section*{Acknowledgments}
\vspace{-3mm}We thank the University of Siegen, 
as well as the Deutsche Forschungsgemeinschaft (DFG, German Research Foundation) through grant 396021762 --- TRR 257 ``Particle Physics Phenomenology after the Higgs Discovery'' 
and through the Exzellenzcluster EXC 3107/1 ``Color meets Flavor -- Suche nach neuen Phänomenen in
der starken und schwachen Wechselwirkung''
for support. 


\vspace{-2mm}



\begin{thebibliography}{99}
\bibitem{Charm2023}\vspace{-3mm}\url{https://indico.physik.uni-siegen.de/event/1/} 
\bibitem{CmF}\url{https://color-meets-flavor.de} 
\bibitem{Instagram}\url{https://instagram.com/subatomic_heroes/}
\bibitem{TikTok}\url{https://www.tiktok.com/@subatomic_heroes}
%\bibitem{Video}\url{https://tp1.physik.uni-siegen.de/wp-content/uploads/sites/6/2025/10/subatomic_heroes_englische_untertitel.mp4}
\bibitem{StrangeCharming}\url{https://web.physik.rwth-aachen.de/user/harlander/CRCPodcasts/}
\bibitem{P3H}\url{https://p3h.particle.kit.edu/}
\bibitem{Spark}\url{https://www.uni-siegen.de/presse/publikationen/spark}
\bibitem{RentProf}\url{https://nt.uni-siegen.de/nt/rent-a-prof}
\bibitem{Masterclass}\url{https://home.cern/tags/masterclass}
\bibitem{TeilchenWelt}\url{https://www.hep.physik.uni-siegen.de/outreach/masterclasses}
%https://www.teilchenwelt.de/}
\bibitem{KinderUni}\url{https://kinderuni-siegen.de/}
\bibitem{CERN70}\url{https://indico.physik.uni-siegen.de/event/390/}
\bibitem{TeilchenZoo}\url{https://teilchenzoo.desy.de/ausstellung/}  
\bibitem{PDG}S. Navas et al.~(Particle Data Group), Phys. Rev. D 110, 030001 (2024) and 2025 update. \url{https://pdg.lbl.gov/} 
\bibitem{Website}\url{https://tp1.physik.uni-siegen.de/outreach}
\end{thebibliography}
\end{document}